\documentstyle[twocolumn,prl,aps]{revtex} 
 
\begin{document} 
\twocolumn[\hsize\textwidth\columnwidth\hsize\csname@twocolumnfalse\endcsname
\draft 
\preprint{} 
\title{Metal-insulator transition in spatially-correlated random 
magnetic field system}
\author{D.N. Sheng, Z. Y. Weng} 
\address{Texas Center for Superconductivity, 
University of Houston, Houston, TX 77204-5506 }  
\maketitle 
\date{today}
\begin{abstract}

We reexamine the problem of delocalization of two-dimensional electrons in 
the presence of random magnetic field. By introducing {\it spatial 
correlations} among random fluxes, a well-defined metal-insulator transition 
characterized by a {\it two-branch scaling} of conductance has been 
demonstrated numerically. Critical conductance
is found non-universal with a value around $e^2/h$. Interesting connections of 
this system with the recently observed $B=0$ two-dimensional metallic phase  
(Kravchenko {\it et al.}, Phys. Rev. B {\bf 50}, 8039 (1994)) are also 
discussed.   
\end{abstract}
]
Whether two-dimensional (2D) electrons can become delocalized in the presence
of random magnetic field (RMF) is still controversial. This is a very 
important issue related to many interesting systems, like half-filled 
quantum Hall effect (QHE) \cite{qhe1,qhe2}, gauge-field description
\cite{high1} of 
high-$T_c$ superconductor and so on. By using the standard transfer-matrix 
method\cite{mac}, a number of numerical calculations 
\cite{numer1,numer2,new} have been performed for a non-interacting 2D 
electron system subject to 
{\it spatially-uncorrelated} RMF. The results indicate that 
electrons are always localized near the band edge, while there is a dramatic
enhancement of localization length as one moves towards the band center.
However, the interpretation of the latter is rather conflicting, ranging
from that all states are still localized\cite{numer1,new} with an
extremely large localization length close to the band center to the
existence of a critical region\cite{numer2} with divergent localization
length. Even if a critical region characterized by wavefunctions with
fractional dimensionality\cite {frac} could exist here, a metallic phase
seems being ruled out by those numerical calculations since a two-branch
scaling as a hallmark for metal-insulator transition (MIT) has never been found.
Analytically, while the study based
on a perturbative nonlinear sigma model approach pointed\cite{theory2} to the 
localization of all states, the existence of extended states was
shown\cite{theory1} possible in the presence of a long-range logarithmic
interaction of the topological density (due to fluctuating 
Hall conductance\cite{cha}), which is
supported by direct numerical calculations\cite{sheng1,yang} of topological
Chern number for the case of spatially-uncorrelated RMF with
{\it reduced} field strength. 

In contrast to spatially-uncorrelated RMF, however, 
magnetic flux fluctuations in realistic systems\cite{qhe1,qhe2,high1} may be  
much more smooth with finite-range spatial correlations. Such a
smoothness can significantly reduce the random scattering effects while still
retain the delocalization effect\cite{theory1,cha,sheng1,yang} introduced by
 magnetic fluxes. In this paper, we demonstrate numerically for the first
 time the existence of MIT which is characterized by a two-branch scaling of
 conductance in the presence of spatially-correlated RMF. The critical
 conductance itself is non-universal, with its value 
around $e^2/h$ which generally increases as the Fermi energy shifts towards 
the band center. With much reduced error bar, the present numerical algorithm
 is also applied to an uncorrelated (white noise limit) RMF case and the 
results unambiguously show that all states are localized without a critical 
region at {\it strong} strength of RMF. Possible connections of the present RMF
 system to the zero-magnetic-field (B=0) 2D metal\cite {krav} are also 
discussed at the end of the paper.

We consider a tight-binding lattice model of noninteracting electrons 
under RMF. The Hamiltonian is defined as follows:
\begin{equation}
H=-\sum_{<ij> } e^{i a_{ij}}c_i^+c_j +\sum _i w_i c^+_i c_i 
\end{equation} 
Here $c_i^+$ is a fermionic creation operator, and $<ij>$ 
refers to two  nearest neighboring sites. 
$w_i$ is an uncorrelated random potential  
(white noise limit) with strength $|w_i|\leq W$. 
 A magnetic flux per plaquette is 
given as  $\phi(k)=\sum _ {\Box} a_{ij}$, where the  summation runs  over four 
links around a plaquette labeled by k. We are interested in the case where
$\phi(k)$ at different k's is correlated which can be  generated in the 
following  way:
\begin{equation}
\phi (k)=\frac{h_0}{\lambda _f^2/4} \sum_i f_i
e^{-|R_{k}-R_{i}|^2/\lambda _f^2}  
\end{equation}
where $R_k$ ($R_i$) denotes the spatial position of a given plaquette $k$($i$). 
$h_0$ and $\lambda _f $ are the characteristic strength and correlation length
scale of RMF, respectively. 
 $f_i $ is a  random  number  distributing uniformly between (-1,+1). 

We employ the following numerical algorithm to calculate the longitudinal 
conductance $G_{xx}$. Based on the Landauer formula, $G_{xx}$ for a square 
sample ${\cal N}=L\times L$ can be determined as a summation over contributions 
from all the Lyapunov exponents of the Hermitian transfer matrix product
$T^+T$\cite{mac,lan}. 
To reduce the boundary
effect of a finite-size system, we connect $M$ different square samples
together to form a very long stripe along $x$ direction [of size 
$L \times (LM)$].  Typically $M$ is chosen to be larger than $5000$ even for
the largest sample size ($L=200$) in this work. In this way, the statistical 
error 
bar is significantly reduced in our results (about $1.5\%$).  
In most of earlier numerical calculations, finite-size localization 
length has been computed where the statistical fluctuation is usually quite
big (especially near the band center) as compared to a direct calculation of
the finite-size longitudinal conductance in the present algorithm.

As a test, we have first re-studied the case in which the flux
$\phi(k)$ is randomly distributed between $-\pi$ to $\pi$ without spatial 
correlations -- the situation investigated previously\cite
{numer1,numer2,new} as
mentioned at the beginning of the paper. We find that $G_{xx} $ monotonically 
decreases with the sample size $L$ at all strengths of the on-site disorders: 
from $W=0$ to $W=4$, and 
is extrapolated to zero at large sample-size limit as shown in Fig. 1 
at a fixed Fermi energy $E_f=-1$. In the insert of
Fig. 1, $G_{xx} $ is shown as a function of the disorder strength $W$ at 
different sample sizes: $L=24$, $80$, and $200$, 
which shows that even at $W=0$ the conductance
monotonically decreases with the increase of $L$, indicating that the dominant 
role of the random flux here is similar to the random potential in causing 
localization of electrons. The one-parameter scaling of $G_{xx} $ can be 
obtained by choosing a scaling variable $\xi$ at each random potential $W$.
As plotted in Fig. 2, all data can be then collapsed onto a single curve of
$L/\xi$, in which $\xi$ is given in the insert of Fig. 2. Clearly
$\xi$ is always finite although it becomes extremely large at weak disorder
limit. This is consistent with the conclusion\cite{numer1,new} that electrons
are all localized and excludes the possibility of a critical 
region\cite{numer2} as the error bar in our calculation is much less than the
variation of the conductance itself.  Notice that in weak-disorder limit 
$\xi $ may no longer be interpreted as localization length\cite{new} which
characterizes an exponential decay of conductance with sample size at strong 
localized region.

Now let us focus on RMF with smooth spatial correlations as defined in (2). 
With the correlation length $\lambda _f =5.0$  
(the lattice constant as the unit) and flux strength $h_0=1$,
 $G_{xx}$ as a function of disorder strength $W$ is computed at a given
Fermi energy $E_f=-1$ as shown in Fig. 3. Curves at different sample sizes 
($L=16$ -- $200$) all cross at a fixed-point $W=W_c $, which is
independent of lattice size $L$ within the statistical error bars.
It is qualitatively different from the behavior of $G_{xx}$ in 
spatially-uncorrelated RMF case discussed above. 
At $W > W_c$, $G_{xx}$ continuously decreases with the increase 
of the sample size, which can be extrapolated to zero at large $L$ limit,
 corresponding to insulating phase. On the other hand, at $W< W_c $, $G_{xx}$
monotonically increases with lattice sizes like a typical metallic behavior.
The insert of Fig. 3 shows the critical conductance $G_{c}$ (corresponding
to $W=W_c$) at different Fermi energies and $h_0$'s. The data of $G_{xx}$ in 
Fig. 3 can be collapsed onto a two-branch curve as a function of scaling 
variable $L/\xi$ as shown in Fig. 4 for $W>W_c$
and $W<W_c$, respectively. The insert of Fig. 4 shows the scaling variable
 $\xi$ vs. $W$ which diverges at the critical point $W_c$. 
In the metallic phase at $W<W_c$, $G_{xx} $  can be  approximately fitted
by the following form: $G_{xx}=G_{s}-c_0*exp(-L/\xi_0)$.  Here $G_s $ is the 
saturated conductance at $L\rightarrow \infty$, 
which is non-universal and depends on the disorder strength $W$ 
as well as  the correlation length $\lambda _f$ of random fluxes.

The introduction of spatial correlations in random fluxes is crucial for such a
metal-insulator transition. We also found a well-defined MIT at an even shorter 
correlation length: $\lambda_f =2.0$. But the larger $\lambda_f$ is, the
stronger the metallic behavior becomes with a larger saturated conductance.
The previously discussed RMF in 
white noise limit may only belong to a very special case in which 
the localization effect of strong randomness of fluxes 
overwrites the delocalization effect of the same fluxes. We would like to
point out that even in such an uncorrelated random flux case, the 
delocalization may be still enhanced if one reduces
the {\it strength} of RMF. Earlier topological Chern 
number calculations\cite{sheng1} clearly indicates a delocalization transition 
as the maximum strength of $\phi(k)$ is reduced to around $\pi/2$. We have
computed the conductance in this case using the present method at much larger 
sample sizes and indeed found a slight increase of the conductance with sample 
size at $W<W_c$, which is  opposite to strong random flux limit where
conductance always decreases with the increase of sample size (Fig. 1),
although a two-branch scaling curves here is not as clear-cut as in
the spatially-correlated RMF case shown in Fig. 4. 
    
As mentioned above, the critical conductance $G_{c}$ varies from $0.5$$ e^2/h 
$ to around $2$$ e^2/h$ as the Fermi energy shifts from the band edge towards 
band center (the insert of Fig. 3). 
It is interesting to note that $G_{c}$ obtained here is in the 
same range as the experimental data found in recent $B=0$ 
2D MIT system\cite{krav}. In the following, we would like to point out a 
possible deeper connection between the two systems. 
 
In a recent experiment\cite{jiang}  in p-type
GaAs/AlGaAs heterostructure, the evolution of delocalized states was studied
 continuously from the  QHE regime at strong magnetic field to zero 
field limit where the $B=0$ MIT is recovered. 
The authors found that the critical density of the lowest extended level in 
QHE regime flattens out, instead of floating up towards infinity, as magnetic 
field is reduced and can be extrapolated to the critical density
of $B=0$ MIT in such a material. Similar result has been also observed in 
Si-MOSFET samples\cite{krav1,sha}. At first sight, it is tempting to think
that the lowest extended level of QHE somehow survives at $B=0$, but 
physically it does not make much sense because QHE extended states carry 
quantized Hall conductance known as Chern number whereas at $B=0$ the total
Hall conductance must be zero without time-reversal symmetry-breaking. In 
fact, experiments indicated\cite{krav1} that before $B$ vanishes, extended
 levels of the QHE may already merge with a different kind of extended level 
(called QHE/Insulator boundary in Ref,\onlinecite{krav1}) which carries an
 opposite sign
of Hall conductance. Theoretically, it has been previously found\cite{sheng2}
that QHE extended states indeed can be mixed with some boundary extended 
level moving down from high-energy side at strong disorder or weak magnetic 
field limit which carries negative Chern number in a lattice model. 
When those extended states with different signs of Chern numbers
mix together at weak magnetic field limit, there could be two consequences:
one is that no states will eventually carry non-zero Chern number due to the
cancellation such that all of them become localized. This is what happens in 
non-interacting system\cite{sheng2}; The second possibility is that individual
states may still carry nonzero Chern numbers and form a delocalized 
{\it region} even though the {\it average} Hall conductance still vanishes at
 $B=0$. Such a system is then physically related to the RMF system where the
 delocalization mechanism is also due to the fluctuating Hall 
conductance\cite{theory1,cha,sheng1,yang}.
Below we give a heuristic argument how a strong Coulomb interaction may lead 
to such a realization. 

At strong Coulomb interaction with $r_s\gg 1 $ (here $r_s $ is the ratio of
the strength of the Coulomb interaction over the Fermi energy\cite {krav}),
the 2D electron state is very close to a Wigner glass phase where the low-lying
spin degrees of freedom may be described by an effective spin Hamiltonian
 $H_s$ given in Ref.\onlinecite{che}. The low-lying charge degrees of freedom
 may be regarded as ``defects''  which can hop on the ``lattice''
governed by a generalized $t-J$ like model\cite{che,voj}. 
Based on many studies on the $t-J$ model in high-$T_c$ problem,
especially the gauge-field description\cite{high1}, charge carriers moving on 
a magnetic spin background can generally acquire fictitious fluxes. Such kind
 of fluxes usually can be treated as random magnetic fields with some 
finite-range spatial correlations. According to the numerical results presented 
above, such a system indeed can have a MIT at $B=0$. Of course, further
model study is needed in order to fully explore this connection 
which is beyond the scope of the present paper.

In conclusion, we have numerically demonstrated the 
existence of a metal-insulator transition characterized by {\it a two-branch 
scaling} for 2D electrons in the presence of {\it spatially-correlated} random 
magnetic fields. In contrast to usual three-dimensional metal where the 
conductance scales to infinity, this 2D metal has a saturated non-universal 
conductance. The range of the critical conductance is very similar to that
found in $B=0$ 2D metal-insulator transition. We briefly discussed a possible 
connection between a 2D interacting electron system at $r_s\gg 1$ and the 
spatially-correlated random-magnetic-field problem based on both experimental 
and theoretical considerations.

{\bf Acknowledgments} -The authors would like to thank  C. S. Ting, X. G. Wen,
and  especially S. V. Kravchenko for stimulating and helpful discussions.
The present work is supported by Texas ARP grant No. 3652707, a grant from 
Robert Welch foundation, and by the State of Texas through the Texas Center 
for Superconductivity at University of Houston.

Fig. 1  The evolution of conductance $G_{xx}$ (in units of $e^2/h$)  with 
sample width $L$ at differnent disorder strength $W$'s for 
spatially-uncorrelated RMF case.  The insert: $G_{xx}$ as a function of
$W$ at different $L$'s.  Fermi energy is fixed at $E_f=-1$.

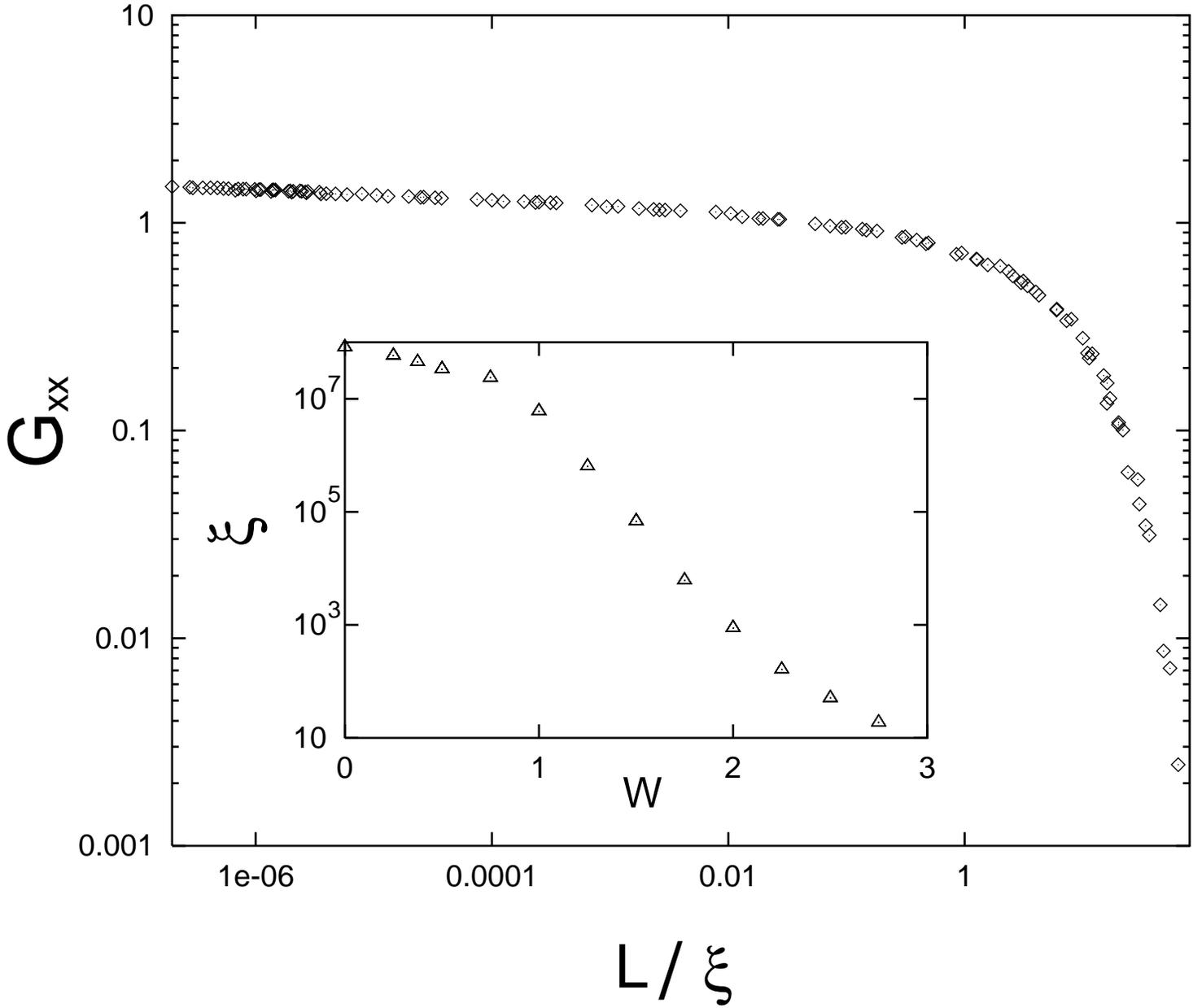
Fig. 2. The data of $G_{xx}$ at  different $L$'s and $W$'s all collapse onto
a scaling curve as a function of $L/\xi$.
The insert: $\xi$  versus $W$.

Fig. 3  $G_{xx}$ versus $W$ at  different sample sizes ($L=16 (\bullet), 
 24,32,48,64,80,120,200 $). $W_c$ is the critical disorder.
Fermi energy is chosen at $E_f=-1$.
The insert: critical conductance $G_{c}$ as a function of Fermi
energy $E_f$.

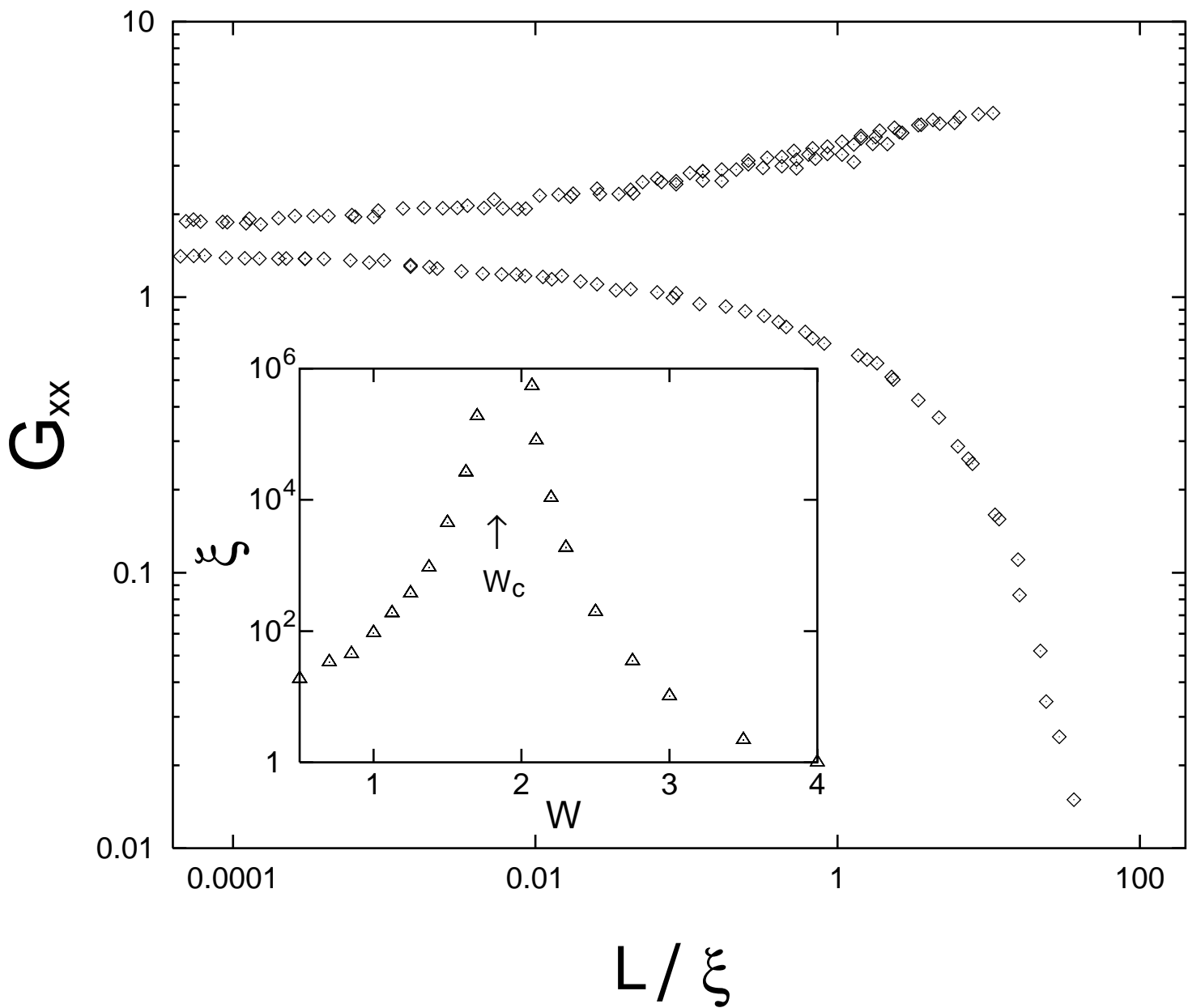
Fig. 4. Two branch-scaling curve of $G_{xx}$ as a single function of 
$L/\xi$  for different  $L$'s and  $W$'s.
The insert: $\xi$  versus   $W$.

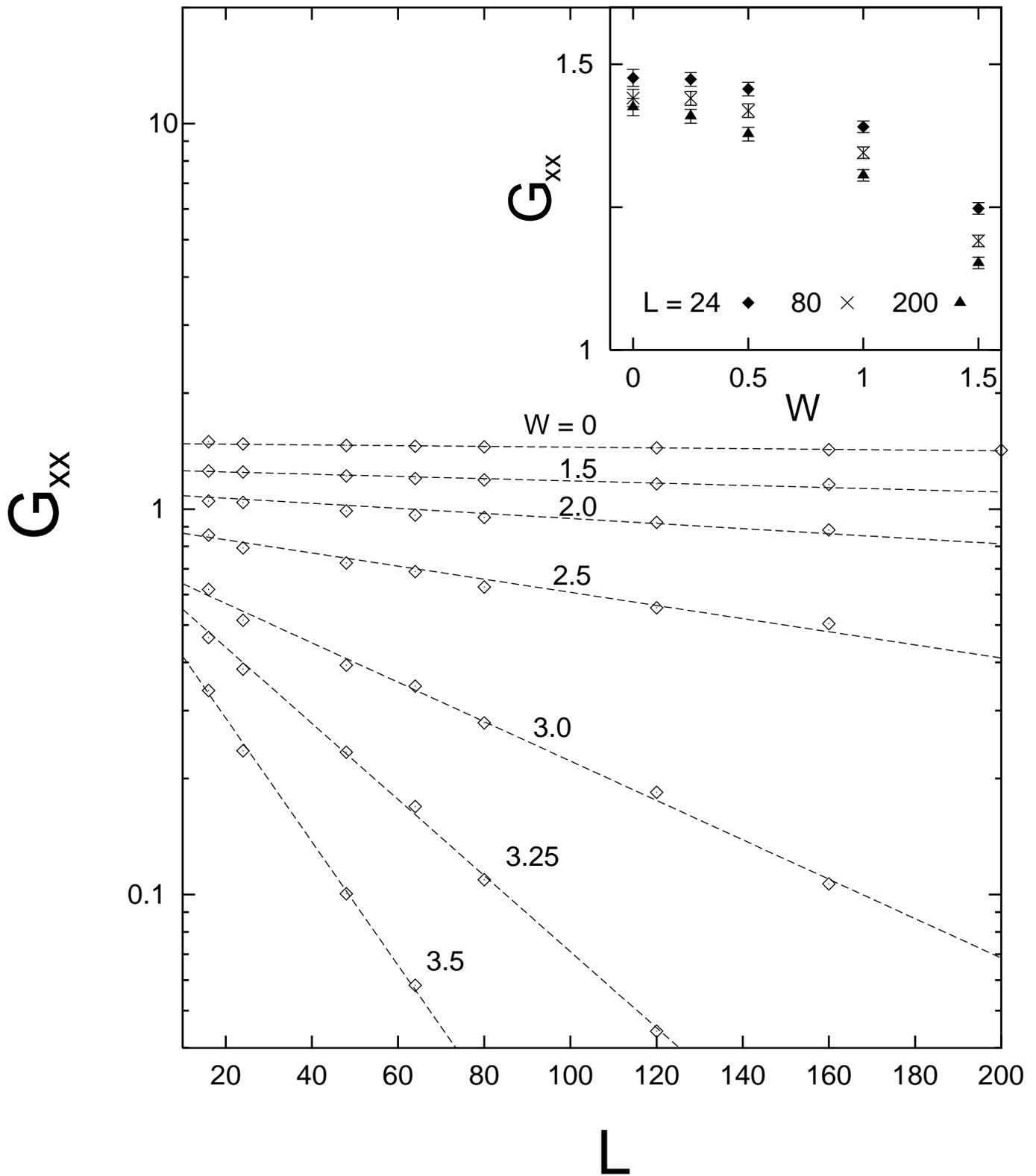
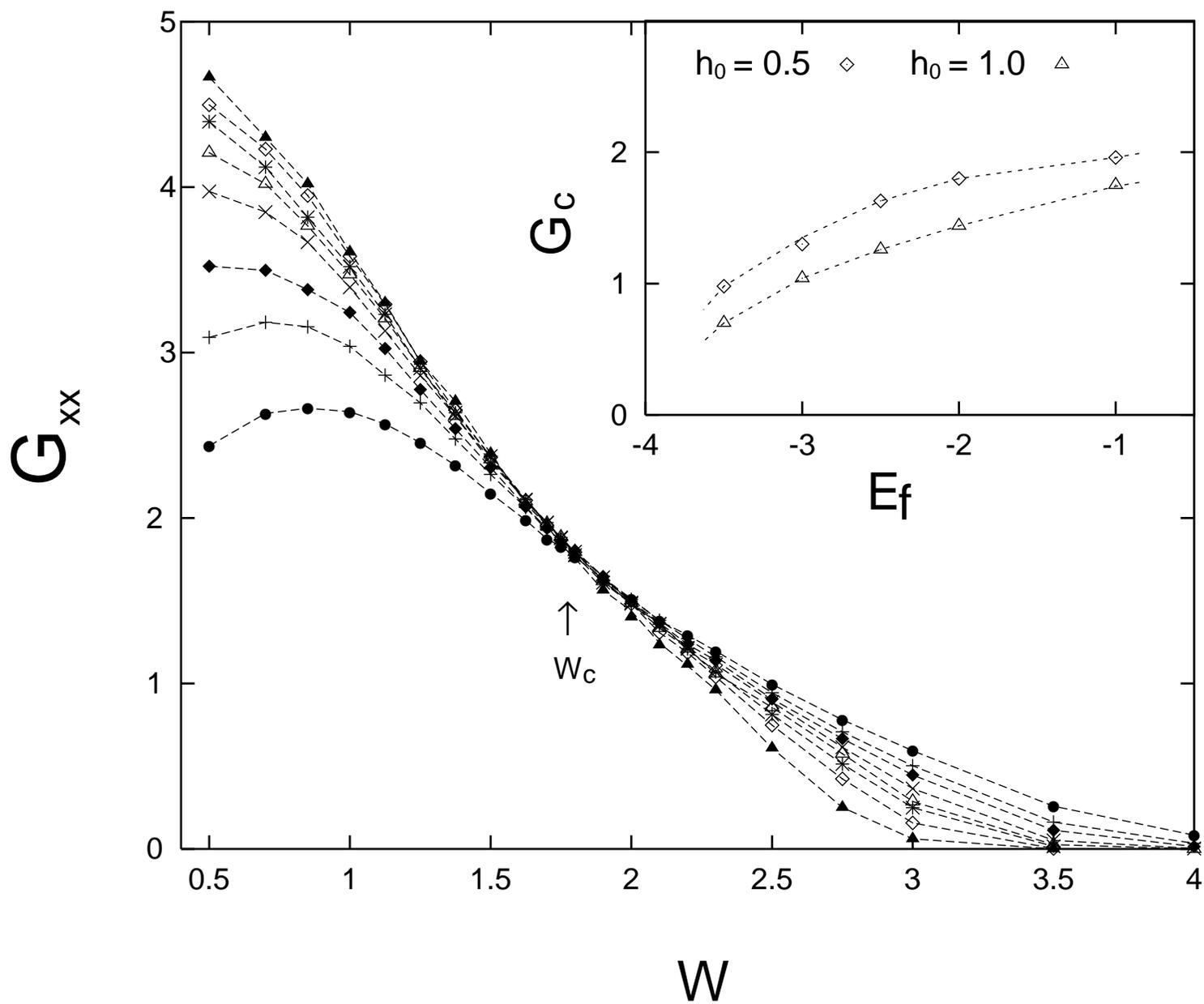
\end{document}